\documentclass[aps,pra,twocolumn,superscriptaddress,nofootinbib]{revtex4-1}
\usepackage{graphicx}
\usepackage{epstopdf}
\usepackage{amssymb}
\usepackage{amsmath}
\usepackage{mathtools}
\usepackage{bm}
\usepackage{xcolor}

\begin{document}

\title{Crossover from weak to strong quench in a spinor Bose-Einstein condensate}

\author{Seji Kang}
\affiliation{Department of Physics and Astronomy, and Institute of Applied Physics, Seoul National University, Seoul 08826, Korea}
\affiliation{Center for Correlated Electron Systems, Institute for Basic Science, Seoul 08826, Korea}

\author{Deokhwa Hong}
\affiliation{Department of Physics and Astronomy, and Institute of Applied Physics, Seoul National University, Seoul 08826, Korea}

\author{Joon Hyun Kim}
\affiliation{Department of Physics and Astronomy, and Institute of Applied Physics, Seoul National University, Seoul 08826, Korea}

\author{Y. Shin}
\email{yishin@snu.ac.kr}
\affiliation{Department of Physics and Astronomy, and Institute of Applied Physics, Seoul National University, Seoul 08826, Korea}
\affiliation{Center for Correlated Electron Systems, Institute for Basic Science, Seoul 08826, Korea}

%\date{\today}

\begin{abstract}
We investigate the early-time dynamics of a quasi-two-dimensional spin-1 antiferromagnetic Bose-Einstein condensate after a sudden quench from the easy-plane to the easy-axis polar phase. The post-quench dynamics shows a crossover behavior as the quench strength $\tilde{q}$ is increased, where $\tilde{q}$ is defined as the ratio of the initial excitation energy per particle to the characteristic spin interaction energy. For a weak quench of $\tilde{q}<1$, long-wavelength spin excitations are dominantly generated, leading to the formation of irregular spin domains. With increasing $\tilde{q}$, the length scale of the initial spin excitations decreases, and we demonstrate that the long-wavelength instability is strongly suppressed for high $\tilde{q}>2$. The observed crossover behavior is found to be consistent with the Bogoliubov description of the dynamic instability of the initial spinor condensate. 
\end{abstract}

\maketitle

\section{Introduction}

Quantum-phase-transition dynamics is a fundamentally important subject concerning how a many-body quantum system evolves into a newly ordered state~\cite{Polkovnikov_rmp,Review,Review2}. Recently, the quantum phase transition of spin-1 Bose-Einstein condensates (BECs) with antiferromagnetic interactions was investigated in ultracold atom experiments~\cite{Bookjans_prl11,Kang_pra17,Vinit_pra17,Kang_prl19}. For zero magnetization, the mean-field ground state of the antiferromagnetic BEC is a polar state with $\langle \hat{\bm{d}}\cdot\mathbf{F}\rangle=0$, where $\hat{\bm{d}}=(d_x,d_y,d_z)$ is a unit spin director and  $\mathbf{F}=(F_x, F_y, F_z)$ is the hyperfine spin operator of the atom~\cite{Ho98,Ohmi98,Kawaguchi_phyrep12}. In an external magnetic field, e.g., along the $z$-axis, a uniaxial spin anisotropy is imposed owing to the quadratic Zeeman energy $E_z=q\langle F_z^2\rangle=q(1-d_z^2)$, and depending on the sign of $q$, two ground states are present in the system: for $q>0$, an easy-axis polar (EAP) state with $\hat{\bm{d}} \parallel \hat{\bm{z}}$ and for $q<0$, an easy-plane polar (EPP) state with $\hat{\bm{d}}\perp\hat{\bm{z}}$. The order parameter manifold of the EAP phase is $U(1)$, whereas that of the EPP phase is $[U(1)\times S^1]/\mathbb{Z}_2$~\cite{Zhou_prl01,Zhou_IJMPB03}. Thus, a quantum phase transition occurs at $q=0$ between the two phases with different symmetries~\cite{Phuc_pra13}.

The quantum-phase-transition dynamics of the antiferromagnetic BEC was experimentally studied using a quantum quench protocol, where the sign of $q$ is suddenly changed~\cite{Bookjans_prl11,Kang_pra17,Vinit_pra17,Kang_prl19}. In our recent work, we investigated the EAP-to-EPP phase transition with  highly oblate, quasi-two-dimensional (quasi-2D) samples~\cite{Kang_pra17}. Spin turbulence was observed to emerge and decay in the quenched BEC, and its time-space scaling properties near the critical point were demonstrated. The creation of half-quantum vortices (HQVs) was observed, resulting from a spontaneous breaking of the continuous symmetry of the spin rotation in the phase-transition dynamics~\cite{Seo_prl15}. We also investigated the backward EPP-to-EAP phase transition by quenching the spin anisotropy in the opposite direction~\cite{Kang_prl19}. Wall-vortex composite defects, which are spin domain walls bounded by HQVs, were observed to be created during the phase transition and their nucleation mechanism was demonstrated.

In this paper, we extend our experiment investigation into the EPP-to-EAP phase transition, with a particular focus on examining how the early-time dynamics changes with increasing the quench strength. The quench strength is quantified as $\tilde{q}=q/(c_2 n)$, with $c_2>0$ being the spin interaction coefficient and $n$ being the atomic density, and thus $\tilde{q}$ represents the ratio of the excitation energy per particle of the initial EPP state to the characteristic spin interaction energy of the system. Our previous study was limited to the defect formation in a weak-quench regime near the critical point~\cite{Kang_prl19}. In this work, we investigate the post-quench dynamics over a wide range of $\tilde{q}$ of up to 4. For a weak quench of $\tilde{q}<1$, the phase-transition dynamics proceeds by generating long-wavelength spin excitations, which is followed by the formation of irregular spin domains and their subsequent relaxation. As $\tilde{q}$ increases, we observe that the characteristic length scale of the initial spin excitations decreases, and for very high $\tilde{q}$, the spin domain structure evolves even into a speckled pattern. Using a sample containing an enhanced seed for long-wavelength spin excitations, we demonstrate that the long-wavelength instability becomes strongly suppressed in a strong-quench regime with $\tilde{q}>2$. We find that the observed crossover of the early-time post-quench dynamics is consistent with the Bogoliubov description of the dynamic instability of the initial EPP state~\cite{Lamacraft_prl,Matuszewski_prl10}. Our results demonstrate the different quantum-quench regimes for a spinor BEC system.

The remainder of this paper is structured as follows. In Sec.~II, we briefly review the Bogoliubov description of the dynamic instability of the initial EPP state after quench. In Sec.~III, we describe our experiment sequence, including the details of the sample preparation. In Sec.~IV, we then present the experiment results, characterizing the crossover behavior of the post-quench dynamics. Finally, some concluding remarks and areas of future study are provided in Sec.~V.

\section{Dynamic instability}

The order parameter of the spin-1 BEC in a polar state can be expressed as 
\begin{equation}
\bm{\psi}=\begin{bmatrix} \psi_{+1} \\ \psi_0 \\ \psi_{-1} \end{bmatrix}= \sqrt{n} e^{i \phi} \begin{bmatrix} -\frac{d_x-id_y}{\sqrt{2}} \\ d_z \\ \frac{d_x+id_y}{\sqrt{2}} \end{bmatrix},
\end{equation}
where $\psi_{j=0,\pm1}$ is the condensate wavefunction of the $m_z=j$ Zeeman component and $\phi$ is the superfluid phase. In the EPP phase with $\bm{d} \perp \hat{\bm{z}}$, i.e. $d_z=0$, the BEC is an equal mixture of the $m_z=\pm 1$ components. According to the Bogoliubov analysis of the EPP state, the system has two branches of magnon excitations~\cite{Kawaguchi_phyrep12,Symes_pra}: gapless axial mode with an energy spectrum of $E_a(k)=\sqrt{\epsilon_k(\epsilon_k+2c_2 n)}$ and gapped transverse mode with $E_t(k)=\sqrt{(\epsilon_k-q)(\epsilon_k-q+2c_2 n)}$, where $\epsilon_k=\hbar^2 k^2/2m$ is the single-particle spectrum with atomic mass $m$. The gapless axial mode is the Goldstone mode associated with the continuous symmetry of the spin rotation in the easy plane. The gapped transverse mode is the excitation mode involving the $m_z=0$ component and its gap energy is given by $\Delta=\sqrt{|q|(|q|+2c_2 n)}>0$ for $q<0$.

The quantum quench of the BEC from the EPP into the EAP phase is driven by changing the sign of $q$ from negative to positive. With $q>0$, the energy $E_t$ of the transverse magnon mode becomes imaginary for a certain range of momentum $k$. This means that, upon the quench, the corresponding magnon modes become unstable and fluctuations in the modes will be exponentially amplified. Consequently, the population of the $m_z=0$ component will grow in the quenched BEC. Its initial growth rate is determined by the maximum magnitude of the imaginary energy as $\Gamma(\tilde{q})\equiv \max \{\text{Im}[2E_t(k)/\hbar]\}$.

In Fig.~1, we show the dispersion curves of the transverse magnon mode, $E_t(k)$, for various values of $\tilde{q}$. For $\tilde{q}<1$, the unstable modes with imaginary energy are restricted to low wavenumbers $k< k_s\sqrt{\tilde{q}}$ [Fig.~1(a)], where $k_s$ is the inverse of the spin healing length $\xi_s=\hbar/\sqrt{2mc_2 n}$. The most unstable mode occurs at zero momentum, giving $\Gamma(\tilde{q})=\gamma_0 \sqrt{\tilde{q}(2-\tilde{q})}$. Here $\gamma_0=2c_2 n /\hbar$ denotes the maximum dynamic instability of the system. When $\tilde{q}$ increases over unity, the momentum of the most unstable mode becomes finite as $k_m=k_s \sqrt{\tilde{q}-1}\neq 0$ and $\Gamma(\tilde{q})$ saturates at $\gamma_0$ [Fig.~1(b)]. Having nonzero $k_m$ implies that competition will occur between the equally unstable magnon modes with the same wavenumber but different  momentum directions. It was anticipated that the quenched BEC will develop qualitatively different spin correlations with a finite $k_m$~\cite{Lamacraft_prl}. When the strength of quench is further increased to $\tilde{q}>2$, the $k$-range of the unstable modes is given by $k_s\sqrt{\tilde{q}-2} < k<k_s\sqrt{\tilde{q}}$ [Fig.~1(c)]. We note that the $k=0$ mode becomes stable for such a high $\tilde{q}$, bringing about a topological change in the unstable mode region in the momentum space. In addition, $k_m$ becomes larger than $\xi_s^{-1}$ and thus the maximally unstable magnon mode has a single-particle-like characteristic. 

\begin{figure}
\centering
\includegraphics[width=8.4cm]{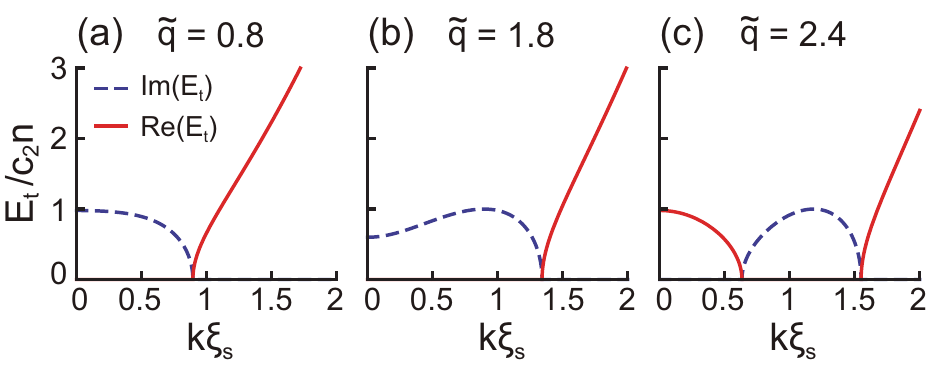}
\caption{Dynamic instability of an antiferromagnetic Bose-Einstein condensate (BEC) quenched from the easy-plane polar (EPP) to the easy-axis polar (EAP) phase. Dispersion curves, $E_t(k)$, of the transverse magnon mode in the quenched BEC for various quench strengths of (a) $\tilde{q}=$ 0.8, (b) 1.8, and (c) 2.4. The solid and dashed lines denote the real and imaginary values of $E_t(k)$, respectively. $c_2 n$ is the spin interaction energy, and $\xi_s$ is the spin healing length of the BEC.}\label{fig:Dispersion}
\end{figure}

\begin{figure*}
	\centering
	\includegraphics[width=16.8cm]{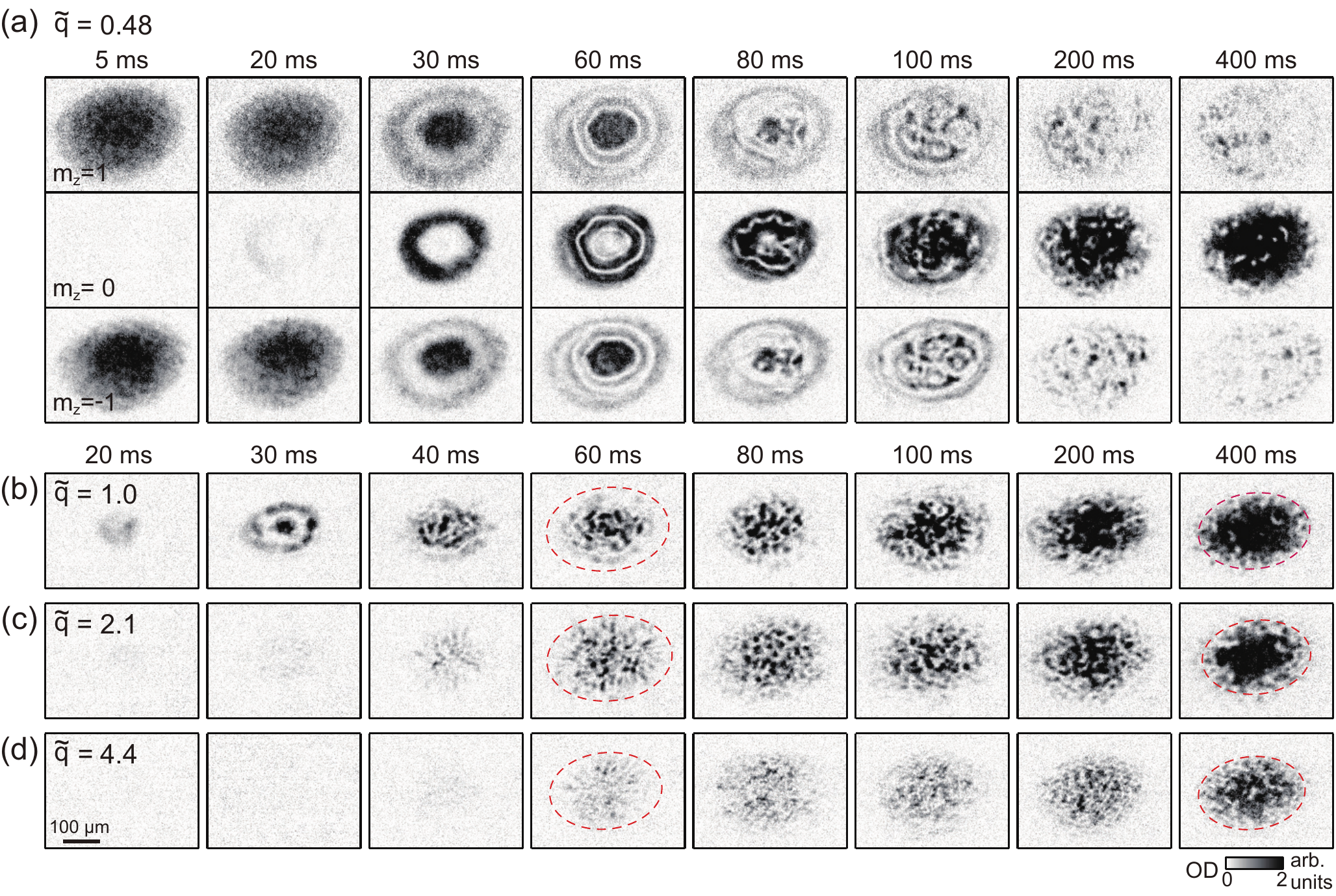}
	\caption{Phase-transition dynamics of an antiferromagnetic BEC from the EPP to the EAP phase. (a) Optical density (OD) images of the $m_z=1,0,-1$ spin components of the quenched condensate for various hold times $t$ after the quench. The quench strength was $\tilde{q}=0.48$, and the images were taken after a 24~ms time of flight with a Stern-Gerlach spin separation. The cloud shapes of the $m_z=\pm1$ components are slightly distorted owing to the inhomogeneity of the spin-separating field gradient. Images of the $m_z=0$ component at various $t$ for (b) $\tilde{q}=1.0$, (c) 2.1, and (d) 4.4. The dashed lines in the images at $t=60$ and 400~ms indicate the boundary of the entire condensate.}\label{fig:EPPtoEAP}
\end{figure*}

\section{Experiment}

Our experiment is performed using a BEC of $^{23}$Na in a $|F=1,m_F=0\rangle$ hyperfine spin state with antiferromagnetic interactions. A highly oblate BEC containing $N_c\approx 5.0\times 10^6$ atoms is prepared in an optical dipole trap (ODT)~\cite{Kang_pra17}. The trapping frequencies of the ODT are $(\omega_x,\omega_y,\omega_z)=2\pi\times(4.3,6.0,440)~$Hz, and the Thomas-Fermi (TF) radii of the trapped condensate are $(R_x, R_y, R_z)\approx(206,148,2.0)~\mu$m. We first prepare a condensate with $\hat{\bm{d}}=\hat{\bm{z}}$ in a magnetic field of $B_z=0.5$~G, and linearly ramp the field down to $B_z=52$~mG in 0.2~s, where the field gradient is controlled at less than 0.1~mG/cm~\cite{Kim_pra19}. We then apply a short rf pulse to rotate the spin direction to $\hat{\bm{d}}=\hat{\bm{y}}$, which transmutes the condensate in the superposition of the $|m_z=+1\rangle$ and $|m_z=-1\rangle$ states, i.e., the EPP state. Immediately after the spin rotation, we tune the quadratic Zeeman energy to $q/h=-5.6$~Hz using a microwave dressing technique and stabilize the EPP state~\cite{Gerbier_pra,Zhao_pra}. We hold the sample under this condition for 0.6~s, during which the fluctuations in axial magnetization are observed to increase to a saturation level~\cite{Seo_prl15}, and we assume that the sample's spin temperature is equilibrated after the holding period. The thermal fraction of the final sample was approximately 15\%.

The EPP-to-EAP phase transition is initiated by suddenly changing the $q$ value to the target value of $q_f>0$. The time evolution of the quenched condensate is probed by taking an absorption image of the sample after a variable hold time of $t$. A Stern-Gerlach (SG) spin separation is applied in the imaging, where, after turning off the microwave field and the ODT, a short pulse of a magnetic field gradient is applied to spatially separate the three Zeeman spin components of the sample during a time of flight of 24~ms.\footnote{During the time of flight, the sample quickly expands along the imaging direction, increasing its thickness to $260~\mu$m. The depth of focus of our imaging system is estimated to be $\approx 100~\mu$m, limiting a full quantitative analysis of the spatial structure of spin domains in this work.} 

In our experiment, the value of $q_f/h$ is controlled within the range of 2.6 -- 82~Hz. The spin interaction energy is $c_2 n_0=h\times 30.7$~Hz for the peak atomic density $n_0$ of the condensate~\cite{Knoop_pra11,Bienaime16,Kim_arXiv19,Black_prl07}.\footnote{$c_2=\frac{4\pi \hbar^2}{m}a_s$
with $a_s=(a_{F=2}-a_{F=0})/3$, where $a_{F=2(0)}$ is the $s$-wave scattering length for a colliding pair of atoms with a total spin of $F=2(0)$. From Ref.~\cite{Knoop_pra11}, $a_s=1.88 ~a_0$, with $a_0$ being the Bohr radius, which was experimentally confirmed in Refs.~\cite{Bienaime16,Kim_arXiv19}. In our previous studies~\cite{Kang_pra17,Kang_prl19}, we used a value of $a_s=0.823 a_0$ from Ref.~\cite{Black_prl07}.} Since the sample thickness $2R_z$ is comparable to the spin healing length $\xi_{s,0}=\hbar/\sqrt{2mc_2 n_0}\approx 2.7~\mu$m, the spin dynamics in the highly oblate sample is effectively 2D, and the magnon dispersion for the 2D spin dynamics is determined by the effective density $\bar{n}=\frac{2}{3}n_0$, which is obtained by averaging the parabolic TF density profile along the tightly confining axial direction under the assumption of hydrodynamic equilibrium~\cite{Stringari98,Zaremba_pra98}. This was experimentally verified in our recent measurement of the speed of spin sound~\cite{Kim_arXiv19}. Calculating the quench strength as $\tilde{q}=q_f/ (c_2 \bar{n})$, the range of $\tilde{q}$ in our experiment is given by $0.12<\tilde{q} <4.4$, covering from the weak to strong regime. In the following, unless specifically mentioned, the related sample parameters are given for the effective density $\bar{n}$, and we have $\xi_s=3.3~\mu$m and $\gamma_0/2\pi = 41.0$~Hz.

\section{Results}

\subsection{Crossover from weak to strong quench}

In Fig.~\ref{fig:EPPtoEAP}(a), we display a sequence of image data for $\tilde{q}=0.48$ to show the time evolution of the quenched BEC in a weak quench regime. Upon the quench, long-wavelength spin excitations are generated in the condensate. According to the Bogoliubov analysis, these correspond to the most unstable transverse magnon modes in the system, and we attribute its ring-shaped spatial structure to the geometry of the trapped condensate~\cite{Klempt_prl09, Scherer_prl10}. As the population of the $m_z=0$ component increases, it is observed that axial polar domains consisting of the $m_z=0$ component are spatially formed. The $m_z=0$ component is immiscible with the $m_z=\pm1$ components for the antiferromagnetic interactions~\cite{Stenger_nat,Jimenez-Garcia_NatCom19}. Because the spin director $\hat{\bm{d}}$ can be either $+\hat{\bm{z}}$ or $-\hat{\bm{z}}$ in the axial polar domains, the domain formation process involves the spontaneous breaking of the $\mathbb{Z}_2$ symmetry, and the domain walls can be constructed at the interfaces of the domains, which are identified in the experiment with their cores occupied by the $m_z=\pm1$ components [see the image at $t=60$~ms in Fig.~2(a)]. In the subsequent evolution, the domain walls are found to be dynamically unstable to split into smaller segments. The image at $t= 80$~ms in Fig.~2(a) shows an undulated domain wall, indicating its snake instability.\footnote{The undulated domain walls were repeatedly observed near the time for the given quench strength.} In our previous work, it was demonstrated that the resultant line segments are composite defects having HQVs at their endpoints~\cite{Kang_prl19}. After the proliferation of the composite defects through the wall-splitting process, the system gradually relaxes into the EAP ground state by depleting the populations of the $m_z=\pm 1$ components, which mainly reside in the core region of the defects.

\begin{figure}
	\centering
	\includegraphics[width=7.8 cm]{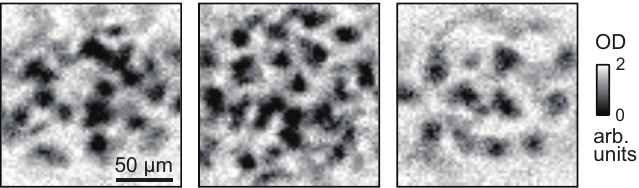}
	\caption{Example images showing an array-like spatial pattern in the density distribution of the $m_z=0$ component. The experiment conditions were $\{\tilde{q},t \}=\{1.0, 50~\text{ms}\}$ (left), $\{2.1, 80~\text{ms}\}$ (middle), and $\{2.1, 60~\text{ms}\}$ (right). The middle image indicates the same data shown in Fig.~1(c). The right image was taken with a sample containing an enhanced seed (see Sec.~IV~C).}\label{fig:Threshold}
\end{figure}

\begin{figure}
	\centering
	\includegraphics[width=8.2cm]{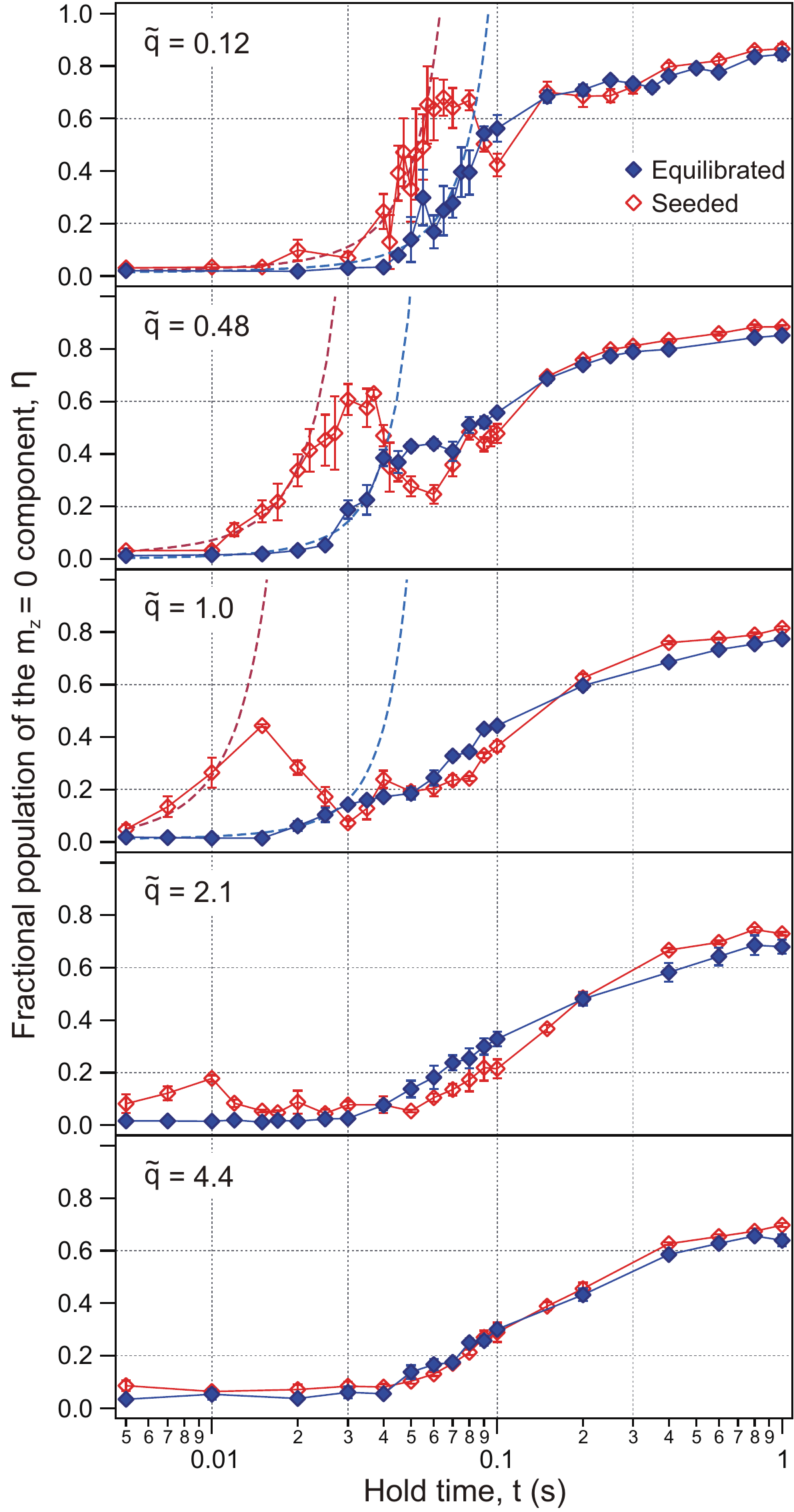}
	\caption{Time evolutions of the fractional population $\eta$ of the $m_z=0$ spin component in the quenched BEC for various values of $\tilde{q}$. The blue solid diamonds indicate the experimental data measured with samples equilibrated in the EPP phase (equilibrated sample), and the red open diamonds indicate those measured with samples that contain a small condensate population of the $m_z=0$ component (seeded sample, see Sec.~IV C). The dashed lines are guidelines for the initial exponential growth of $\eta$. Each data point was obtained by averaging at least four measurements of the same experiment, and the error bar indicates the standard error of the mean value. 
 }\label{fig:dynamics}
\end{figure}

In Figs.~2(b)--2(d), we display additional image data sets for three different higher values of $\tilde{q}$. As the quench strength increases, we observe that the spatial structure of the incipient spin domains formed by the $m_z=0$ component becomes finer, which indicates that the characteristic length scale of the initial spin excitations is shortened. This observation is consistent with the prediction that the momentum range of the unstable modes increases with increasing $\tilde{q}$. It is interesting to note that when $\tilde{q}$ increases over unity, we occasionally observe that the incipient spin domains of the $m_z=0$ state appear granulated, forming an array-like pattern [see the images for $t=80$~ms in Figs.~2(b)and 2(c)]. A few examples are provided in Fig.~3. In the Bogoliubov analysis of the initial dynamic instability, the wavenumber of the most unstable mode is predicted to be finite for $\tilde{q}>1$ as $k_m=k_s\sqrt{\tilde{q}-1}$. Thus, if the post-quench evolution is dominantly driven by multiple magnon modes having finite wavenumbers with different directions, it can give rise to the formation of a quasi-periodic spin structure~\cite{Matuszewski_prl10}.

When the strength of quench further increases to $\tilde{q}>2$, we observe that the initial spin structure evolves into a speckled pattern [Fig.~2(d)]. In this high-$\tilde{q}$ case, the domain-wall formation is not clearly identified because of the small domain size. This suggests that the picture of the composite defect nucleation obtained in the weak-quench regime might not be applicable to the high-$\tilde{q}$ case. The domain-wall structure should be significantly modified for high $\tilde{q}\gg1$ because it is energetically too costly to host the $m_z=\pm1$ components in the core region~\cite{Takeuchi}.

\subsection{Time evolution of spin composition}

We characterize the phase-transition dynamics by measuring the time evolution of the fractional population $\eta(t)$ of the $m_z=0$ spin component (Fig.~\ref{fig:dynamics}). Here $\eta=N_0/N_c$ and $N_c=N_{+1}+N_0+N_{-1}$, where $N_{j=0,\pm1}$ is the $m_z=j$ atom number of the condensate and is determined from the SG spin-separation absorption imaging. The value $\eta$ indicates the fractional energy released into the system from the initial quadratic Zeeman energy, and the post-quench dynamics is the system evolution from $\eta=0$ to $\eta=1$. Initially, $\eta$ shows an exponential growth owing to the amplification of the unstable magnon mode. When $\eta$ increases to a certain threshold of $\eta_{th}$, its growth behavior becomes gradual. We can see that, at around this point, the domain walls start forming in the quenched condensate, and the change indicates that the system enters a late-time stage where it undergoes coarsening and relaxation dynamics of the spin domains. This two-step evolution of the spin composition was also observed in the EAP-to-EPP transition dynamics~\cite{Kang_pra17,Symes_pra18}.

The growth curve of $\eta$ is characterized using two parameters, $\{t_1, t_2\}$, which are the times at which $\eta=0.2$ and 0.6, respectively. In Fig.~5 we plot the measurement results of $t_1$ and $t_2$ as a function of $\tilde{q}$. The time $t_1$ represents the onset time of the $m_z=0$ component, which can be compared with the inverse of the instability rate, $\Gamma^{-1}$, assuming that the initial magnitude of the spin fluctuations in the corresponding unstable modes is not significantly dependent on $\tilde{q}$. In the weak-quench region of $\tilde{q}<1$, the measured $t_1$ is observed to decrease, following the decreasing behavior of $\Gamma^{-1}\propto \tilde{q}^{-1/2}$. However, as $\tilde{q}$ increases over unity, $t_1$ increases, deviating from $\Gamma^{-1}$, which settles to its minimum value. We find that the deviation is due the decrease of $\eta_{th}$ below 0.2 for high $\tilde{q}>1$, which means that the initial exponential growth stage is already finished before $t_1$. Here, $\eta_{th}$ monotonically decreases from $\approx0.6$ to $\approx 0.15$, with $\tilde{q}$ increasing to unity (Fig.~4). 

\begin{figure}
	\centering
	\includegraphics[width=7.6cm]{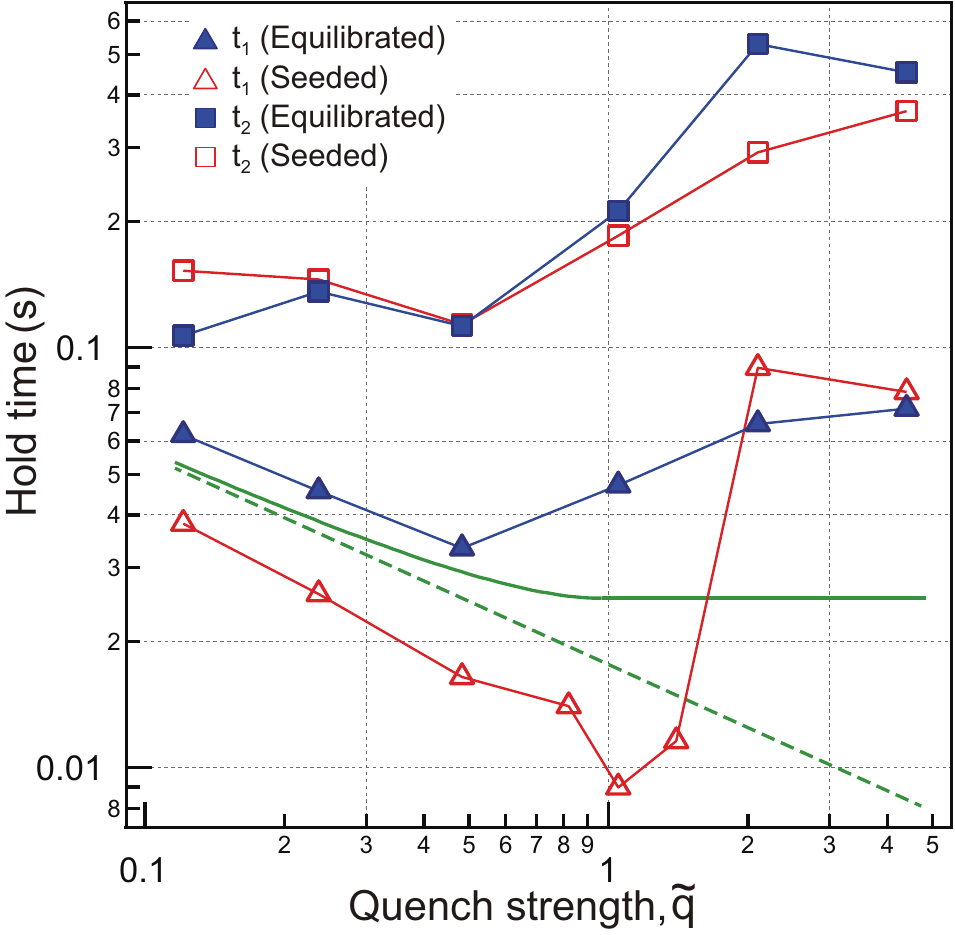}
	\caption{Characterization of the growth curve of $\eta$ as a function of the quench strength $\tilde{q}$. Here, $t_1$ (triangles) and $t_2$ (squares) are the hold times after quench when $\eta$ increases to 0.2 and 0.6, respectively. The solid green line denotes $\Gamma^{-1}(\tilde{q})$, free from a proportional factor, and the dashed green line is a guideline for $\tilde{q}^{-1/2}$ scaling.}\label{fig:Power}
\end{figure}

The time $t_2$ at which $\eta=0.6$ is observed to increase with increasing $\tilde{q}$ (Fig.~5). This indicates that the relative relaxation speed of the spin domains becomes slower with their finer spatial structure, which might be understood as the saturation effect of the energy dissipation rate. In the strong-quench regime with $\tilde{q}>2$, the dominant spin excitation mode is predicted to have a single-particle-like characteristic with $k_m>\xi_s^{-1}$, and hence the subsequent spin-domain formation and coarsening dynamics will be qualitatively different from those in the weak-quench regime. The late-time relaxation dynamics of the spin turbulence deserves further investigation~\cite{Symes_pra18,Fujimoto_pra122,Karl_scirep,Witkowska_pra14,Fujimoto_pra16,Williamson_prl16,Schmied_arXiv19}.

\subsection{Quench dynamics with an enhanced seed}

In the evolution of the dynamic instability of the quenched condensate as $\tilde{q}$ is increased, a key characteristic is that the long-wavelength instability disappears for high $\tilde{q}>2$. To check this, we conduct the same quench experiment with a sample containing a small condensate population of the $m_z=0$ component. Since the small addition of the $m_z=0$ component can be regarded as zero-momentum transverse magnons, the participation of long-wavelength spin excitations will be selectively enhanced in the post-quench evolution of the sample. We prepare such a sample by removing the spin-temperature equlibrating stage during the sample preparation sequence, i.e., without the holding period in the EPP phase after the rf pulse. In our experiment, the rf pulse duration was optimized to minimize the population of the $m_z=0$ component, but as shown in the following, its residual population immediately after the coherent spin rotation is much larger than that in the sample equilibrated in the EPP phase. 

\begin{figure}
\centering
	\includegraphics[width=8.4cm]{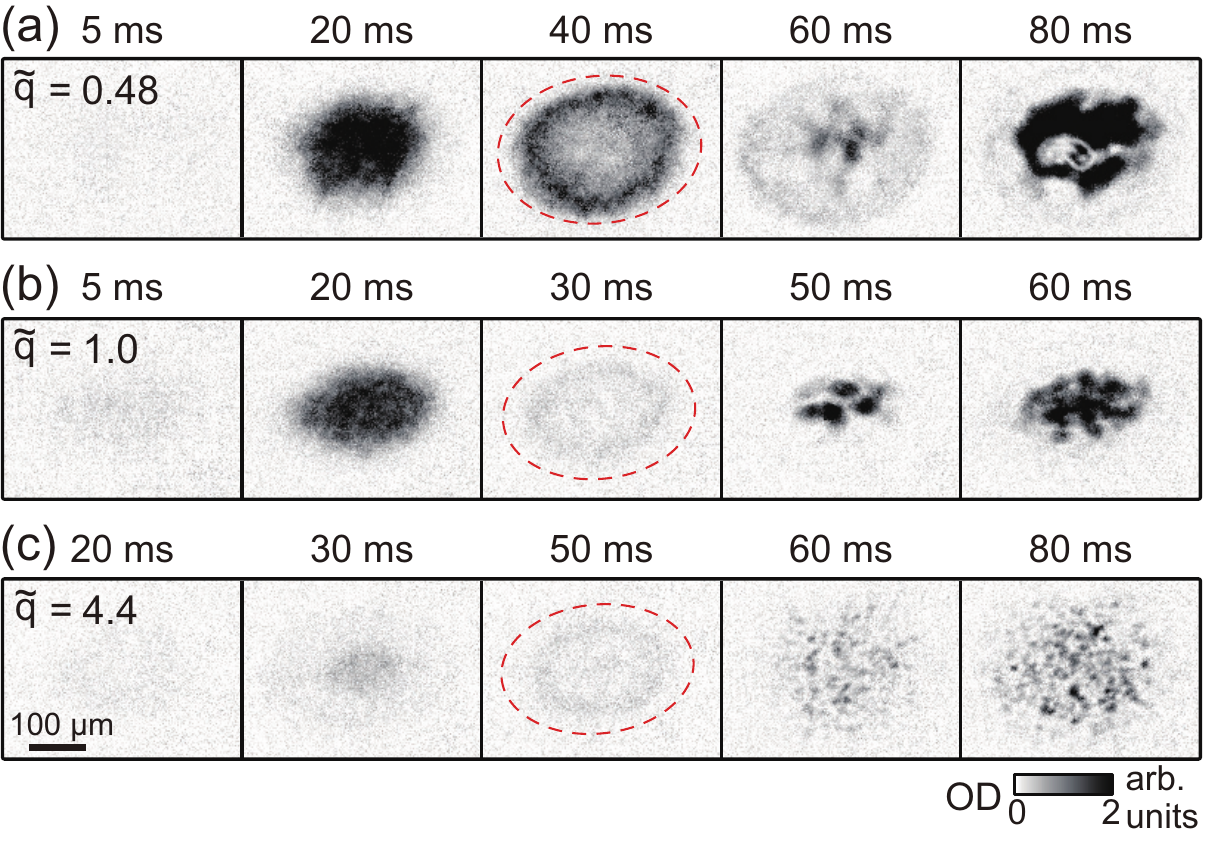}
	\caption{Images of the $m_z=0$ component at various hold times $t$ for (a) $\tilde{q}=$ 0.48, (b) 1.0, and (c) 4.4. The sample was prepared to contain a small condensate population of the $m_z=0$ component, corresponding to the seeded sample in Figs.~4 and 5.}\label{fig:EPPtoEAP2}
\end{figure}

In Fig.~6 we show images of the $m_z=0$ component at short hold times of $t\leq 80$~ms to demonstrate the time evolution of the seeded sample after quench. In a low-$\tilde{q}$ case, long-wavelength spin excitations are generated and spin domain formation follows, as observed in the previous experiment using the equilibrated sample. However, it should be noted that with the seeded sample, the generation of spin excitations occurs much faster in time, and furthermore, it is more coherent in that the population of the $m_z=0$ component oscillates with a large amplitude. The faster onset time and the collective spin oscillation clearly indicate that the initial population of low-momentum transverse magnons is selectively enhanced in the seeded sample.

The growth curve of $\eta$ for the seeded sample is displayed in Fig. 4, along with that for the previous equilibrated sample. During the early stage of the evolution, $\eta$ exhibits a temporal pulse, corresponding to the collective spin oscillation, and then returns to its original growth curve. We observe that the amplitude of the collective spin oscillation monotonically decreases with increasing $\tilde{q}$ and nearly vanishes for $\tilde{q}>2$. For the highest $\tilde{q}\approx 4$, the growth curve of $\eta$ is identical to that of the original sample, demonstrating that the role of the long-wavelength spin excitations in the post-quench dynamics largely fades within the strong-quench regime. This observation is consistent with the prediction that the $k=0$ magnon mode becomes stable for $\tilde{q}>2$.

In Fig.~5, the onset time $t_1$ is plotted as a function of $\tilde{q}$. The $\tilde{q}$-dependence of $t_1$ shows good agreement with the predicted instability of $\textrm{Im}[E_t(k=0)] \propto \sqrt{\tilde{q}(2-\tilde{q})}$ for $\tilde{q}<2$. In our experiment, the shortest $t_1$ was measured to be $\approx 10$~ms at $\tilde{q}\approx 1$, which is approximately 4 times shorter than the shortest onset time for the equilibrated sample. When we model the initial growth of $\eta$ as $\eta(t)=\alpha e^{\gamma_0 t}$ with the maximum growth rate of $\gamma_0$, the fourfold decrease of $t_1$ implicates that the seed population of the low-$k$ magnon mode in the seeded sample is $\approx 10^4$ times larger than that in the equilibrated sample. It was noted that the quantum quench can be used as the magnon thermometry of a spinor BEC~\cite{Mele_pra13}.

\section{Summary and outlook}

We investigated the early-time dynamics of an antiferromagnetic spin-1 BEC after a sudden quench from the EPP to the EAP phase and observed its characteristic evolution with increase in the quench strength. Our experiment results show three distinctive quench regimes: weak, intermediate, and strong. In the weak-quench regime, the post-quench dynamics is driven by long-wavelength spin excitations; in the intermediate regime, it is mainly driven by spin excitations with finite wave numbers; and in the strong regime, the role of long-wavelength spin excitations is strongly suppressed. We found that the crossover of the post-quench dynamics is consistent with the evolution of the dynamic instability predicted from the Bogoliubov analysis of the quenched condensate.

All characteristics of the crossover originate from the magnon dispersion of the initial EPP phase. Because the EAP phase has the same dispersion structure of two degenerate magnon modes as $E(k)=\sqrt{(\epsilon_k +q)(\epsilon_k+q+2c_2n)}$, we expect that the early-time dynamics of the EAP-to-EPP transition will show the same crossover behavior as the initial excitation energy is increased~\cite{Kang_pra17}, although the subsequent late-time dynamics will differ owing to the different symmetry of the ground state.

A natural extension of this study will be to investigate, with better image resolution, the spin domain formation dynamics as a function of $\tilde{q}$. As mentioned before, the domain-wall structure will be significantly modified with a high $\tilde{q}$ so the defect nucleation mechanism will be qualitatively changed when $\tilde{q}$ increases into a strong regime. In addition, the late-time relaxation dynamics of the spin turbulence and its possible scaling behavior are interesting subjects to explore in future experiments~\cite{Fujimoto_pra122,Symes_pra18,Karl_scirep,Witkowska_pra14,Fujimoto_pra16,Williamson_prl16,Schmied_arXiv19}.

\begin{acknowledgments}
We thank H. Takeuchi for discussion and critical reading of the manuscript. This work was supported by the Samsung Science and Technology Foundation (Project No. SSTF-BA1601-06).
\end{acknowledgments}

\end{document}